\pdfoutput=1
\documentclass[reprint,aip,apl, twocolumn,floatfix,superscriptaddress]{revtex4-1}
\usepackage{graphicx}
\usepackage{dcolumn}
\usepackage{bm}
\usepackage{amsmath}
\usepackage{lineno}
\usepackage[english]{babel}
\usepackage[utf8]{inputenc}
\usepackage{color}
\usepackage{siunitx}
\usepackage{amsmath}
\usepackage{textcomp, gensymb}
\usepackage{upgreek}
\usepackage{hyperref}
\hypersetup{
    colorlinks=true,
    citecolor=magenta,
    linkcolor=blue,
    filecolor=magenta,      
    urlcolor=magenta,
    }

\begin{document}

\title{Conductance Plateaus at Quantum Hall Integer Filling Factors in Germanium Quantum Point Contacts}

\author{Karina Hudson}
\affiliation{QuTech and Kavli Institute of Nanoscience, Delft University of Technology, Lorentzweg 1, 2628 CJ Delft, Netherlands}
\author{Davide Costa}
\affiliation{QuTech and Kavli Institute of Nanoscience, Delft University of Technology, Lorentzweg 1, 2628 CJ Delft, Netherlands}
\author{Davide Degli Esposti}
\affiliation{QuTech and Kavli Institute of Nanoscience, Delft University of Technology, Lorentzweg 1, 2628 CJ Delft, Netherlands}
\author{Lucas E. A. Stehouwer}
\affiliation{QuTech and Kavli Institute of Nanoscience, Delft University of Technology, Lorentzweg 1, 2628 CJ Delft, Netherlands}
\author{Giordano Scappucci}
\email{g.scappucci@tudelft.nl}
\affiliation{QuTech and Kavli Institute of Nanoscience, Delft University of Technology, Lorentzweg 1, 2628 CJ Delft, Netherlands}

\date{\today}

\begin{abstract}
Constricting transport through a one-dimensional quantum point contact in the quantum Hall regime enables gate-tunable selection of the edge modes propagating between voltage probe electrodes. Here we investigate the quantum Hall effect in a quantum point contact fabricated on low disorder strained germanium quantum wells. For increasing magnetic field, we observe Zeeman spin-split 1D ballistic hole transport evolving to integer quantum Hall states, with well-defined quantised conductance increasing in multiples of $e^2/h$ down to the first integer filling factor $\nu=1$. These results establish strained germanium as a viable platform for complex experiments probing many-body states and quantum phase transitions.
\end{abstract}

\maketitle

Quantum point contacts (QPCs) offer a versatile platform for investigating many-body correlations in low-dimensional condensed matter systems. A well-known case is the so-called $0.7$ anomaly, a shoulder-like feature in the conductance occurring at $G=0.7\times2e^2/h$, which has been interpreted as arising from the formation of a quasi-bound state in the constriction exhibiting Kondo or Kondo-like correlations \cite{CronenwettPRL02,MeirPRL02}. Beyond this, QPCs can be coupled to mesoscopic reservoirs, such as metallic islands, enabling experimental access to the multichannel Kondo regime \cite{IftikharNat15,IftikharSci18}. Furthermore, QPCs have also be used to tune the conduction of many-body fractional quantum Hall edge modes\cite{NakamuraPRL23} and to select 2D edge modes for creating coupled Kondo states, of interest for quantum simulation of quantum phase transitions\cite{PouseNP23}.

Thus far, the majority of experimental studies of many-body physics and quantum Hall edge states in hole quantum point contacts have been reported in high-mobility GaAs heterostructures \cite{RokhinsonPRL06,DanneauPRL08, KlochanPRL11,KomijaniPRB13,NakamuraPRL23}. Amongst group IV semiconductor, strained germanium quantum wells\cite{ScappucciNatRevMat21} are an alternative promising platform due to the high mobility\cite{StehouwerAPL23, MyronovSS23, CostaAPL24} and light effective mass\cite{LodariPRB19}. However, germanium remains comparatively unexplored in the context of QPCs, despite the rapidly growing importance as a host material for low-noise spin qubits in quantum dots\cite{StehouwerNat25}, superconductor–semiconductor devices for hybrid quantum systems\cite{TosatoNatComms23} and other fundamental physics phenomena due to its strong spin–orbit coupling. Ballistic 1D hole transport has previously been reported in germanium-based QPCs, as well as anisotropic Zeeman spin-splitting of the 1D conductance subbands in the out-of-plane field direction due to a large out-of-plane $g$-factor\cite{GulAPL17, MizokuchiNL18, GaoArXiV24}. However, gate-tunable transmission of integer Landau edge modes in germanium QPCs has not been previously reported.

In this Letter, we study high-field magnetotransport in QPCs fabricated on a strained germanium quantum well~\cite{LodariMatQT21}, which was used previously in several spin qubit experiments~\cite{HendrickxNatComms20,HendrickxNat21, WangSci24, ZhangNatNano25}. In the absence of a magnetic field we observe characteristic quantised ballistic transport in the QPC. In a small out-of-plane magnetic field we observe Zeeman spin-splitting of the 1D subbands along with an upward shift in energy due to the the field coupling to the hole orbital momentum. When the field is sufficiently large such that the radius of cyclotron orbit is on a similar or smaller length scale than the width of the QPC, we observe a transition to 2D quantum Hall physics and the emergence of integer Landau levels, with evidence for well-defined quantum Hall plateaus down to the first $\nu =1$ Landau level filling.

\begin{figure*}[ht]
    \centering
    \includegraphics[width=\linewidth]{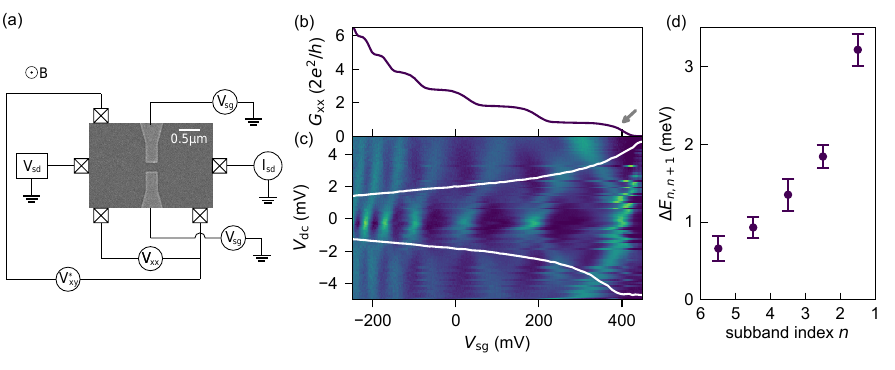}
    \caption{(a) Scanning electron micrograph of a nominally identical quantum point contact (QPC) device and schematic of the circuit. An ac-voltage signal $V_{\mathrm{sd}}$ is applied across the QPC, and $I_{\mathrm{sd}}$, $V_{\mathrm{xx}}$, and $V^*_{\mathrm{xy}}$ are measured.
    (b) Conductance $G$ as a function of split-gate voltage $V_{\mathrm{sg}}$. Plateaus form at integer multiples of $2e^2/h$ indicating ballistic 1D conductance inside the QPC constriction. The trace has beencorrected for a series resistance $R_{\mathrm{s}}=\SI{2.4}{\kilo\ohm}$). The grey arrow indicates the $0.7\times 2e^2/h$ anomaly.
    (c) Colour map of source-drain bias transconductance $\partial G/\partial V_{\mathrm{sg}}$ (a.u.) plotted against $V_{\mathrm{dc}}$ and $V_{\mathrm{sg}}$. Light blue regions correspond to risers in 1D conductance and dark blue regions correspond to plateaus. The white overlaid traces indicate the dc voltage drop across the QPC.
    (d) 1D subband spacing $\Delta E_{n,n+1}$ as a function of subband index $n$ extracted from source-drain bias spectroscopy in panel (c).}
    \label{fig1}
\end{figure*}

The quantum point contact device is a two-gate layer stack fabricated on undoped accumulation mode Ge/SiGe quantum well heterostructure on a silicon wafer with peak mobility $\mu = 2.5\times 10^5 \SI{}{\centi\meter\squared\per\volt\per\second}$\cite{LodariMatQT21} using standard electron beam lithography techniques\cite{LawrieAPL20}. The two-dimensional hole gas (2DHG) forms in the compressively-strained Ge quantum well, positioned $\SI{55}{\nano\meter}$ below the dielectric interface, via a global top gate and has been characterised in previous work\cite{LodariMatQT21}.
The 2DHG is further confined using a pair of QPC split gates $\SI{300}{\nano\meter}$ long and $\SI{300}{\nano\meter}$ wide to deplete the 2DHG and create a ballistic 1D channel between two 2DHG reservoirs. A scanning electron micrograph of the device along with a schematic of the measurement circuit is provided in Fig.~\ref{fig1}(a). Measurements were performed in a dilution refrigerator with a base temperature of $\SI{85}{\milli\kelvin}$ at the mixing chamber, using low-frequency ac lock-in techniques with an applied excitation voltage $V_{\mathrm{sd,ac}}=\SI{100}{\micro\volt}$. Measurements presented here were performed at a fixed 2D hole density of $p=1.6 \times 10^{11} \SI{}{\per\centi\meter\squared}$, calculated from the frequency of Shubnikov-de Haas oscillations when the QPC gates are open in the 2D regime. The longitudinal voltage $V_{\mathrm{xx}}$ was measured across the sample, while the Hall voltage $V^*_{\mathrm{xy}}$ is measured diagonally across the QPC. 

As a preliminary characterisation, we show in Fig.~\ref{fig1}(b) longitudinal conductance at $B=\SI{0}{\tesla}$ calculated as $G_{\mathrm{xx}}=\frac{h}{2e^2}/\left(\frac{V_{\mathrm{xx}}}{I_{\mathrm{sd}}} - R_{\mathrm{s}}\right)$ where $R_{\mathrm{s}}$ is the series resistance correction accounting for the resistivity of the 2D hole reservoirs between the QPC constriction and the voltage probe contacts. The conductance exhibits quantised plateaus in integer multiples of $2e^2/h$ as a function of side gate voltage $V_{\mathrm{sg}}$ characteristic of ballistic 1D transport. The shoulder-like structure at $0.7\times 2e^2/h$ is an expected feature in QPC conductance and can be attributed to the formation of a quasi-bound state in the vicinity of the saddle-point potential formed by the QPC gates\cite{CronenwettPRL02, SloggettEPJB08, BauerNat13}. Moving on to finite bias spectroscopy, Fig.~\ref{fig1}(c) is a colour map of transconductance $dG/dV_{\mathrm{sg}}$ as a function of $V_{\mathrm{sg}}$ and $V_{\mathrm{sd,dc}}$. 
Applying a dc voltage causes the 1D subbands to evolve from conductance quantised at multiples of $2e^2/h$ to odd multiples of $e^2/h$, from which the subband spacing $\Delta E_{n,n+1} =eV_{\mathrm{dc}}$ can be extracted and is plotted with respect to subband index $n$ in Fig.~\ref{fig1}(d)\cite{PatelPRB91}. The values found here for the energy spacing between 1D subbands are consistent with those reported in a similar strongly confined split-gate QPC with a global top gate architecture in Ref~\cite{GaoArXiV24}. The 1D subband energy spacing in this germanium hole QPC is an order of magnitude larger than the 1D subband spacing of holes in comparable QPC devices in GaAs~\cite{DanneauAPL06,DanneauPRL08}. The larger 1D subband spacing of holes in germanium can be attributed to germanium having a lighter hole effective mass $m^*_h \approx 0.06m_e$\cite{LodariPRB19} compared to $m^*_h \approx 0.4m_e$ for holes in GaAs\cite{BroidoPRB85}. The energy spacings of the germanium 1D hole subbands reported in Ref~\cite{MizokuchiNL18} are an order of magnitude smaller, which is likely due to the different electrostatic confinement profile created from using an architecture consisting of a channel gate and two side gates.

\begin{figure}
    \centering
        \includegraphics[width=1\linewidth]{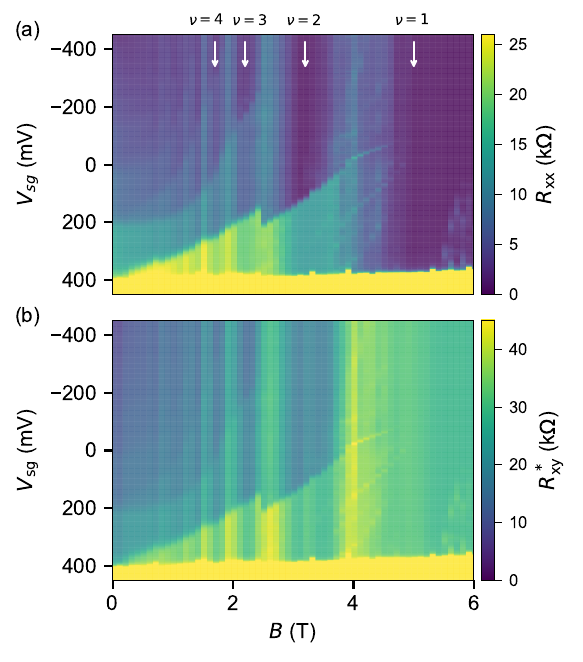}
    \caption{(a) Colour map of longitudinal resistivity $R_{\mathrm{xx}}$ as a function of split gate voltage $V_{\mathrm{sg}}$ and out-of-plane magnetic field $B_{\perp}$. The bottom of the map is the region of QPC pinch off; the 1D subband occupation of the QPC approaches the 2D definition point near the top of the map. The integer Landau filling factors are labelled.
    (b) Colour map of diagonal resistivity $R^*_{\mathrm{xy}}$ as a function of split gate voltage $V_{\mathrm{sg}}$ and out-of-plane magnetic field $B_{\perp}$. }
    \label{fig2}
\end{figure}

Figure ~\ref{fig2} (a) is a colour map of longitudinal resistance $R_{\mathrm{xx}}$ plotted against split-gate voltage $V_{\mathrm{sg}}$ and out-of-plane perpendicular magnetic field $B_{\perp}$. The lighter regions correspond to high resistance. Darker regions correspond to low or zero resistivity. 
On the far left side of the panel at $B=0$: the 1D subband occupation is zero in the vicinity of QPC pinch-off at $V_{\mathrm{sg}} > \SI{400}{\milli\volt}$. 
As the $V_{\mathrm{sg}}$ becomes more negative, the electrostatic confinement of the QPC widens, and 1D subbands occupy the constriction. 
The region where $R_{\mathrm{xx}} \approx \SI{12.9}{\kilo\ohm}$ at $400>V_{\mathrm{sg}}>200$ corresponds to the first 1D conductance plateau $2e^2/h$. As $V_{\mathrm{sg}}$ becomes more negative, additional 1D subbands occupy the QPC and $R_{\mathrm{xx}}$ decreases in steps. 
Where $V_{\mathrm{sg}}<\SI{-400}{\milli\tesla}$, the QPC gates are no longer electrostatically confining the holes to a 1D channel and the system is effectively 2D. 
As $B$ increases towards $\SI{1}{\tesla}$, the subbands undergo Zeeman spin splitting. 
The subbands also have a noticeable upward curvature due to $B_{\perp}$ coupling with the orbital momentum of the holes, which further confines the holes and pushes the subbands higher in energy. 
When $B$ is sufficiently large such that the magnetic length $l_B$ of a given 1D subband is smaller than the width of the QPC channel, integer Landau levels start to emerge. 
At $B\gtrsim \SI{4.5}{\tesla}$, the $n=1$ subband evolves into the $\nu=1$ Landau level and the system is purely 2D. 
At $B=\SI{5}{\tesla}$, $R_{\mathrm{xx}} =0$ and only a single edge mode is propagating through the QPC. At $B=\SI{3.2}{\tesla}$ when $V_{\mathrm{sg}}<\SI{200}{\milli\volt}$, $R_{\mathrm{xx}} =0$ and two edge modes are propagating through the QPC; when $\SI{400}{\milli\volt}>V_{\mathrm{sg}}>\SI{200}{\milli\volt}$, a single spin-resolved ballistic conductance channel occupies the QPC.

Figure ~\ref{fig2} (b) is a colour map of Hall resistance $R^*_{\mathrm{xy}}$ plotted against split-gate voltage $V_{\mathrm{sg}}$ and magnetic field $B_{\perp}$. Similarly to Fig. ~\ref{fig2} (a), the lighter regions correspond to high resistivity, and darker regions correspond to low resistivity. The 1D Zeeman spin-split subbands are visible at $B<\SI{1}{\tesla}$ and Landau levels emerge beyond $B>\SI{1.5}{\tesla}$. The fine fringe-like structures that appear in both $R_{\mathrm{xx}}$ and $R_{\mathrm{xy}}$ at $B=\SI{4}{\tesla}$ can be attributed to tunnelling between the 1D ballistic channel and the 2D $\nu=1$ Landau level\cite{SmolinerPRB91}. Beyond $B\approx \SI{5.5}{\tesla}$, field-induced disorder begins to emerge and we are unable to observe fractional states expected at higher field in this measurement.
The field-induced disorder emerges first in the 1D quantum limit in the vicinity of $V_{\mathrm{sg}}=\SI{400}{\milli\volt}$ and can be interpreted as the magnetic length approaching the scale of fluctuations in the 2DHG potential~\cite{EfrosSSC89}. 

\begin{figure}
    \centering
    \includegraphics[width=\linewidth]{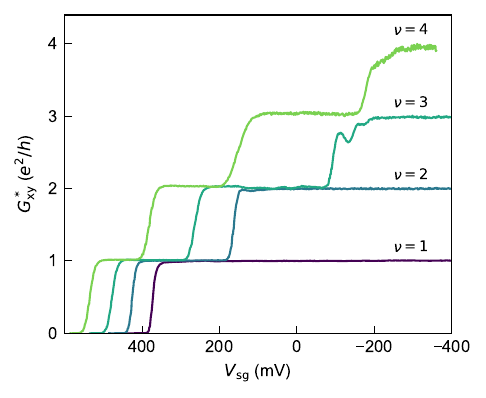}
    \caption{Transverse (Hall) conductance $G^*_{\mathrm{xy}}$ as a function of split gate voltage $V_{\mathrm{sg}}$ at Landau filling factors $\nu = 1,2,3,4$, taken at $B = $\SI{5}{\tesla}, \SI{3.2}{\tesla}, \SI{2.2}{\tesla}, \SI{1.7}{\tesla} respectively. Each trace has been corrected for a fixed series resistance of $R_{\mathrm{s}} = \SI{5.6}{\kilo\ohm}, \SI{3.3}{\kilo\ohm}, \SI{4.9}{\kilo\ohm}, \SI{6.7}{\kilo\ohm}$, respectively. Traces have been offset in $V_{\mathrm{sg}}$ for clarity.}
    \label{fig3}
\end{figure}

Figure ~\ref{fig3} shows the conductance $G$ with respect to $V_{\mathrm{sg}}$ at the $\nu=1,2,3,4$ Landau levels, taken at $B = 5.0,\,3.2,\,2.2,\,1.7\si{\tesla}$ respectively. These line cuts correspond to the indicated regions of $R_{\mathrm{xx}}$ and $R_{\mathrm{xy}}$ in Fig.~\ref{fig2} and have been corrected for the series resistance $R_{\mathrm{s}}$ of the 2DHG reservoirs (which varies with respect to the out-of-plane field) and offset in $V_{\mathrm{sg}}$ for clarity~\cite{repo2025}. In the absence of a magnetic field the 1D conductance shows up to six clear plateaus at integer multiples of $2e^2/h$ (see Fig. ~\ref{fig1}b). A $\nu=1$ the conductance exhibits only a single plateau at $G = e^2/h$. As the Landau level occupation of the QPC constriction increases to $\nu=2$, then $\nu=3$, and additional edge modes are able to propagate, and conductance steps emerge at $G=2e^2/h$ and $G=3e^2/h$ respectively. All four line-cuts exhibit well-defined steep pinch-offs indicating that there is no disorder such as spurious charge traps in the vicinity of the 1D constriction. Small oscillations on the plateau of $\nu=2$ and more substantially on the plateau of $\nu=3$ are likely Fabry-Perot oscillations due to interference of reflected edge modes\cite{HeyderPRB15}. 

The excellent quantisation particularly at filling factors $\nu=1$ and $\nu=2$ demonstrates that strained germanium is a viable platform for implementing more complex experiments probing many-body states and quantum phase transitions typically performed on III-V materials~\cite{PouseNP23}.
This is despite strained germanium quantum wells for this study were grown on a silicon substrate, having known structural disorder due to the lattice mismatch and fluctuating strain-related defects impacting on 2D hole gas mobility\cite{LodariMatQT21,StehouwerAPL23}. More recent generations of strained Ge quantum wells grown on germanium substrates are substantially less disordered due to one order of magnitude less dislocations with an order of magnitude higher mobility and $2\times$ smaller percolation density~\cite{StehouwerAPL23}. We anticipate that future QPC devices fabricated on such wafers offer a potential path forward to explore more complex quasiparticle phenomena such as fractional quantum Hall effects, and chiral edge states\cite{NakamuraPRL23, LaubscherPRB24}.

\bigskip

\small{We thank David Goldhaber-Gordon and Praveen Srirama for their helpful insights and feedback. We acknowledge support by the European Union through the IGNITE project with grant agreement No. 101069515 and the QLSI project with grant agreement No. 951852.
This work was supported by the Netherlands Organisation for Scientific Research (NWO/OCW), via the Open Competition Domain Science - M program. We acknowledges the research programme Materials for the Quantum Age (QuMat) for financial support. This programme (registration no. 024.005.006) is part of the Gravitation programme financed by the Dutch Ministry of Education, Culture and Science (OCW). This research was sponsored in part by The Netherlands Ministry of Defence under Awards No. QuBits R23/009. The views, conclusions, and recommendations contained in this document are those of the authors and are not necessarily endorsed nor should they be interpreted as representing the official policies, either expressed or implied, of The Netherlands Ministry of Defence. The Netherlands Ministry of Defence is authorized to reproduce and distribute reprints for Government purposes notwithstanding any copyright notation herein.}

\subsection*{Authors contributions and declarations}
L.E.A.S. grew the Ge/SiGe heterostructure developed with input from G.S. K.L.H. fabricated the QPC device with input from D.D.E., performed measurements with input from D.C. and wrote the manuscript with input from G.S.

 G.S. is founding advisor of Groove Quantum BV and declares equity interests.

\subsection*{Data availability statement}

The data sets supporting the findings of this study are
openly available at the Zenodo repository~Ref.~\cite{repo2025}.

\end{document}